**Title**: How to Do STEM Outreach Evaluation – Recommendations Based on a Review of Self-Evaluation Tools in Canadian STEM Outreach Programs

**Authors**: Garrett Richards (Environmental Policy Institute, grichards@mun.ca) and Svetlana Barkanova (Physics, sbarkanova@mun.ca) of the School of Science and the Environment, Grenfell Campus, Memorial University of Newfoundland and Labrador

**Abstract**: STEM (Science, Technology, Engineering, and Mathematics) outreach programs in Canada, especially those oriented towards youth, play a critical role in supporting the nation's future workforce, innovation capacity, and equity across social groups in STEM fields. They constitute a large, multi-layered ecosystem connecting universities, national laboratories, non-profit organizations, and grassroots community groups. Despite the growing importance of these programs, and the frequency that they undergo self-evaluation, little systematic information exists on best practices or common approaches to evaluating the effectiveness of STEM outreach initiatives. To address this gap, we integrated literature review with email inquiries about self-evaluation tools sent to Canadian STEM outreach programs funded by NSERC (Natural Sciences and Engineering Research Council of Canada) PromoScience grants. We contacted 200 programs and heard back from about 100 of them, for a response rate of 50%. Of those respondents, 68 shared information about a formal self-evaluation tool appropriate for general STEM outreach. The results led us to develop a toolbox of self-evaluation methods, master question banks and starting-point templates for student/participant and teacher/chaperone surveys, and a synthesized list of recommendations for evaluation process, design, and implementation. Our approach provides a broad treatment of how to do STEM outreach evaluation, supplementing the relevant literature, where large-N studies and Canadian studies are relatively rare. We acknowledge that some of the most effective practices in STEM outreach evaluation require resources or capacity (e.g. longitudinal approaches), which may be limited for many outreach practitioners, but others seem to have a high ratio of benefit to cost (e.g. adding qualitative questions to an otherwise quantitative survey).

**Acknowledgements**: We are grateful for the support of research assistants for this project; many thanks to Natasha Pennell (literature review, data collection, and data summary), Ali Mujahid (literature review and data collection), and Edward Asante (data analysis). We are also grateful for peer feedback from: Eden Hennessy on the template questions and surveys in the appendices—and Japna Sidhu-Brar and Kevin Hewitt on the collected literature.

## Introduction

Exposure to STEM outreach sparks curiosity and encourages young people to consider careers they might not have otherwise imagined, especially if they are members of groups underrepresented in STEM fields. STEM outreach also plays a crucial role in fostering an innovative, inclusive, and scientifically literate (McGinnis, 2020) society—and it is one way that STEM practitioners can fulfill a broader social responsibility to engage the public about their work. Fortunately, STEM practitioners in Canada offer a rich landscape of outreach programs, such as school visits, summer camps, and hosted tours. Outreach evaluation is an important dimension of this landscape, because it can help to maximize the quality and impact of the programs. However, clear and detailed guidance on "how to do effective STEM outreach evaluation" is not readily available for STEM practitioners.

This paper is a collaboration between a social scientist (first author) and a natural scientist (second author). The former—with an interest in science-policy interfaces and science communication—was conducting a study interviewing STEM outreach practitioners from academia and from government, in comparison. In doing so, he met the latter—with an interest in physics (especially astronomy) outreach to diverse groups—who was developing evaluation methods for her own outreach program for girls, Indigenous students, and rural youth in Newfoundland, Canada. A physicist with minimal experience in social science research, she mentioned seeking guidance from the relevant literature but encountered challenges in finding accessible and applicable practices. Working together, with the support of research assistants, the two authors explored the literature more systematically, which indeed contained very few pieces with clear and thorough answers on "how to do effective STEM outreach evaluation" and was also lacking in large-N studies and Canadian studies, although piecemeal lessons can be extracted. The literature also revealed that outreach evaluation, while rather useful in nearly any form, may tend towards inconsistency, misalignment with program goals, and lack of rigour (see next section for elaboration).

Thus, the purpose of this paper is to build knowledge on how to do effective STEM outreach evaluation, which includes identifying and developing tools as well as delineating considerations for their use. We pursue this purpose using comprehensive literature review, a survey of Canadian STEM outreach programs, qualitative analysis, and reflection. Following this introduction, the second section elaborates the literature review, the third section details methods for the survey of nearly 100 Canadian STEM outreach programs, the fourth section explains the principal results of the survey (with a bank of questions included in appendices), the fifth section discusses key themes emerging from the results, and the sixth section synthesizes recommendations for STEM outreach practitioners by bringing together the survey discussion with the literature (with template outreach evaluation questionnaires included in appendices).

## Background

The pre-existing literature contains several insights on the question of how to do STEM outreach evaluation. We found that such literature can be divided into three categories. First, most commonly, there are pieces reporting specific outreach activities, including evaluation such as a participant survey, but with no broader discussion of that evaluation (e.g. the evaluation was presented mainly as evidence of activity success). Second, there are pieces that similarly report outreach activities, but include a brief or incidental discussion of outreach evaluation more broadly (e.g. rationale, design, or limits of evaluation). Third, and most rarely, there are pieces where outreach evaluation is a central focus of discussion. In this section, we identify relevant themes in the literature, with focus weighted toward the rarer, more informative pieces.

One foundational theme is the general importance of outreach evaluation, best exemplified by Varner (2014), whose framework for evidence-based outreach argues that evaluation is a key component of effective outreach activities. They propose seven steps (emphasis ours): define goals, collaborate for design, tailor to a specific audience, implement the activity, *formative assessment*, *evaluate and reflect*, and *share results*. Similarly, Haywood & Bessley (2014, p. 99) state that robust evaluation strategies start by identifying relevant indicators that are clear and justified, while Bussard (2021) advises that the metrics of success vary across different objectives and audiences. In short, doing good STEM outreach requires thinking about goals, and evaluation for measuring progress towards those goals, from the beginning. The inclusion of outreach evaluation results in many other pieces also implicitly reinforces the importance of evaluation.

Limitations of outreach evaluation are discussed extensively in the literature. One fundamental challenge is the tendency for evaluation measurements to be misaligned with program goals. Indeed, Sadler et al. (2018) and Tran et al. (2024, p. 28) found that, in most cases, outreach evaluations privilege measuring delivery (e.g. enjoyment of participants) rather than measuring progress towards underlying goals (e.g. more people working in STEM fields). Of course, long-term goals (e.g. more people working STEM fields) are difficult to measure due to impracticality and complex cause-effect relationships (Laursen et al., 2007; Henderson et al., 2015). The short-term priorities and constrained capacity of universities and funding agencies also limit support for more intensive evaluations (Conference Board of Canada, 2020; Sadler et al., 2018; Tran et al. 2024, p. 37).

Another challenge for outreach evaluation is the tendency to lack rigour. The most common type of evaluation is self-reported measures of participant interest (Franklin et al., 2013, p. 371). Such measurements are not linked to more favourable perception of STEM fields broadly or greater interest in STEM careers (Tran et al., 2024, p. 7)—they are more subjective and less useful than revealed or observed measures (Clark et al., 2016, pp. 6-7; Sami et al., 2020, p. 24), such as comparing pre- and post-surveys on content (e.g. Sami et al., 2020) or beliefs (e.g. Innes et al., 2013). The use of experimentation to compare treatment and control groups (e.g. Tsybulsky et al., 2017) or pseudo-experimentation to compare treatment groups and census data (e.g. Constan & Spicer, 2015) are particularly rigorous, though impractical in many contexts.

A similar shortcoming, related to the misalignment of (long-term) goals and (short-term) measurements, is that most outreach evaluations take only one "on the day" measurement, immediately after the outreach activity (Sadler et al., 2018). Such approaches, regardless of the questions asked, only truly measure how much participants enjoyed the event, rather than any lasting impact on belief, learning, or intention (Laursen et al. 2007, p. 50). A longitudinal component—such as repeat quizzes for the same participants over a multi-week initiative (e.g. Okcay et al., 2021) or interviewing teachers for observations of their students several weeks after attending an outreach event (Laursen et al., 2007)—provides better evidence of impact, because it can measure whether learning or attitude changes persist past the short term. For example, Dewilde et al. (2019) reported significant learning retained five months after their outreach event, although the effect had diminished compared to immediately after the event. Pieces reporting specific outreach activities fall into five categories based on how they attend to long-term impact: no consideration of long-term impact, one-time measures attempting to project potential long-term impacts (e.g. asking "are you interested to be a scientist?" before and after the activity as in Kerby et al. 2010), short-term follow-ups to check for lasting impacts weeks or months later (e.g. clicker surveys administered to repeat attendees to the same event as in Dewilde et al. 2019), medium-term follow-ups a few years later (e.g. interviews with program alumni as in Constan & Spicer, 2015), and true long-term follow-ups after many years (an unrealistic possibility discussed hypothetically in many pieces).

Overall, many outreach practitioners may not be aware of how to conduct evaluation rigorously, instead relying on anecdotal experience (Bagiya, 2016; Laursen et al., 2017). The extensive study of Bagiya (2016) concludes by identifying a desperate need in the field, calling for a "generic evaluation tool" to be developed. Comparably, the review by Tran et al. (2024, p.

36) concludes with a call for more consistent instruments and “guidance for how outreach evaluators ought to select and validate instruments”. However, even one-time, self-reported measures can still be useful; Niemann et al. (2004) found that certain aspects of their outreach program were reported as less enjoyable than others and were able to alter the program’s design to rectify the difference. These measures are also perfectly appropriate for programs with goals beyond long-term career interest in STEM fields, particularly those serving remote or under-resourced communities. For example, consider: providing short-term enjoyment, building a general positive association with STEM fields, informing students of broad career options (Conference Board of Canada, 2020, p. 3).

What options for outreach evaluation are available, to address the above limitations for instance? Sadler et al. (2018, p. 594) observe five categories of tools used across outreach programs at nine case universities: pre-surveys immediately before the event, observation of participants during delivery, post-surveys immediately after the event (most common by far), interviews with facilitators, and longitudinal surveys or other data about past or repeat participants. They also pointed out the informal measure of whether schools make repeat bookings. Comparably, the Conference Board of Canada (2020, pp. 17-19) reports five comparable categories used in Canadian outreach programs for Indigenous students: immediate post-event feedback (i.e. surveys, conversations, or interviews), instructor observations and reflections, activity metrics (e.g. number of participants reached, workshops run, communities engaged, program hours delivered), positive outcome indicators (e.g. participant interest maintained through multiple days of the event, repeat requests from teachers, event projects successfully completed by participants), and longer-term impacts (as measured by testimonials and success stories). Note the range of time scales from immediate to long-term and the range of evaluation participants across participants, facilitators, and teachers.

As for specific questions asked within various evaluation measures, several pieces of STEM outreach literature helpfully include participant-oriented evaluation tools in their entirety (e.g. in a table, photo, or appendix). Dubetz and Wilson (2013, p. 44) and Islam et al. (2018, p. 41) share examples of typical short surveys, Okcay et al. (2021, p. 16) share an example of a pre-post-quiz model emphasizing content, and Sami et al. (2020, p. 28) share an example of a pre-post survey focusing on knowledge around a particular skillset. From these examples, general questions that could be used for any outreach event include varying versions of: “how much did you enjoy this event?”, “are you more interested in STEM after this event?”, “did this event match your expectations?”, “what was your favorite part of the event?”, and “what improvement would you suggest for next time?” (in addition to non-evaluative questions like demographic questions and logistical questions such as how they found out about the event).

Some of the more intensive outreach studies (e.g. Niemann et al., 2004; Tsybulsky, 2019; Vennix et al., 2018) list a wide range of questions, akin to databases, covering topics such as: learning professional skills, whether the event was too technical, attitudes towards STEM fields, whether the participant felt anxious, and motivations for STEM interest. Tran et al. (2024, p. 27) lists 18 such instruments, highlighting a tool called S-STEM (see Unfried et al. 2015) as an example of one with rigorous validation. Examples of tools and questions are a useful starting point for considering how to do evaluation, but most of the pieces that share these examples do not explicitly discuss evaluation design (i.e. how to decide what tools or questions to use) and the comprehensive instruments therein, while extensive, are often more about STEM attitude or intention and not about outreach directly.

Explicit discussions of how to do STEM outreach evaluation are very rare in the STEM outreach literature. The best example is Bagiya's (2016) extensive study involving interviews with outreach practitioners and teachers, out of which emerge several clear findings on evaluation design: quantitative survey questions are not as useful as qualitative survey questions, ongoing relationships between practitioners and teachers can lead to longer-term insights, and approaches to evaluation should be shared and discussed across practitioners. Tran et al. (2024) also pay attention to the methodology of outreach evaluation—whether quantitative, qualitative, or mixed—concluding that qualitative components tend to facilitate stronger analysis. Another specific recommendation comes from the Conference Board of Canada (2020, p. 21): evaluation of outreach programs for Indigenous students should be thoughtfully planned in advance, in collaboration with Indigenous communities, to ensure that the intended beneficiaries of the program are truly considered (i.e. as opposed to solely external funder priorities or research goals). Finally, the Australian Government's (n.d.) National STEM Education Resources Toolkit has a thorough section titled "I want to evaluate a STEM initiative" that provides a high-level framework for thinking about outreach evaluation (e.g. start by determining the purpose and deciding who should be involved, then build and customize your evaluation framework with the appropriate tools), though it is not specifically focused on outreach and does not get into the details of individual measurement tools.

A few other potential lessons emerge from incidental discussions of evaluation design in other relevant pieces. Keep the evaluation short and simple if the event is fast-paced (Kerby et al., 2010, p. 1026). Use 10 or fewer questions so that it is easy for participants to provide feedback (Yawson et al., 2016, p. A121). Acknowledge the potential biases of question order (i.e. where encountering one question may influence the participant's response to a subsequent question) and thus consider multiple versions of a survey to measure and mitigate such biases (Dewilde et al., 2019, pp. 2665-2666). Consider prioritizing formative (i.e. mid-point) evaluations that may trigger immediate improvements in addition to summative (i.e. end-point) evaluations (Varner, 2014, p. 337). One common cross-cutting theme is the importance of stakeholder partnerships in designing outreach activities and evaluation (Bagiya, 2016; Conference Board of Canada, 2020; Henderson et al., 2015; Varner, 2014); collaboration at the outset ensures that the evaluation will be valuable to the event's intended beneficiary, while continued collaboration after the event opens up broader possibilities for evaluation. There are also design considerations suggested implicitly by the literature on evaluation limitations discussed above: base evaluations on evidence not just anecdotal experience, measure progress toward program goals not just program delivery, use revealed or observed measures not just self-reported measures, and measure longer-term outcomes not just immediate results. Table 1 presents a summary of general lessons for STEM outreach evaluation design.

Table 1. Summary of General Lessons for Effective Design of STEM Outreach Evaluation from the Relevant Literature

| |
|---|
| • set goals for the evaluation process (and the outreach initiative in general) proactively |
| • if possible, work with appropriate stakeholders to design the evaluation from the start (especially if equity-deserving groups are a target audience) |
| • consider formative (mid-point) evaluations as well as summative (end-point) evaluations |
| • prioritize evidence over anecdotal experience in evaluation design |
| • keep the evaluation short and simple, especially if the event is fast-paced |
| • surveys should use 10 or fewer questions |
| • include qualitative questions (and consider prioritizing them over quantitative questions) |
| • acknowledge and counter the biases of question order |
| • consider revealed measures over self-reported measures |
| • measure progress towards program goals; if those goals are longer-term outcomes, appropriate longer-term measures should follow the initial evaluation, if capacity allows |

This brief literature review reveals that the STEM outreach literature is scarce in: explicit and comprehensive discussions on how to do outreach evaluation, large-N studies of many outreach initiatives at once, and Canadian studies. Our research responds to each of these gaps by surveying the evaluation tools used in Canadian STEM outreach programs. The literature review also reveals considerations for our analysis. For example, will Canadian outreach evaluations show the same tendency towards surveys, self-reported measures, more focus on program delivery than program goals, short-term measurements, and under-use of evidence in evaluation design?

## Methods

To achieve our purpose of understanding and improving self-evaluation practices in Canadian STEM outreach programs, we identified English-language NSERC PromoScience programs as cases for study (acknowledging that STEM outreach happens in countless other contexts as well). NSERC is one of Canada's three main research funding agencies, along with SSHRC (Social Sciences and Humanities Research Council) and CIHR (Canadian Institutes of Health Research). NSERC PromoScience is a non-research grant program that funds post-secondary institutions, non-profit organizations, museums, and science centers to conduct STEM outreach to elementary and high school students over 1-3 years (NSERC, 2024). Successful recipients of PromoScience grants are listed on the NSERC website, including links to the funded initiative (which typically include contact information), serving as a publicly available list of many outreach programs in Canada. The PromoScience program requests, but does not require, post-event information on: total number of participants engaged, whether programs were designed to reach equity-deserving groups (i.e. girls, Indigenous youth, visible minorities, youth from low-income backgrounds, youth with disabilities, youth from rural/remote areas), % of participants with increased interest in STEM, % of participants with increased understanding of STEM, % of participations with plan to (continue to) study STEM, and % of participants interested in a STEM career.

In the summer of 2020, we combed the public listings for all PromoScience grants awarded between 2015 (when PromoScience started) and 2018 (the most recent year for which information was available at the time). We were able to locate contact information for 199 English-language programs and sent a short email message to each one, asking whether they used any sort of self-evaluation method and, if so, whether they would be willing to share that method with us. We heard back from 96 programs (i.e. a response rate of 48.2%), of which 68 used some sort of self-evaluation tool appropriate for general STEM outreach evaluation. Most respondents shared the tool in its entirety (e.g. a file containing a survey they had used), but some simply described their evaluation process.

To make sense of the dataset, we used a qualitative analysis process over two iterations of categorization. First, we identified a range of categories around the type of evaluation (e.g. participant survey, quiz, teacher/chaperone surveys) and applied labels to each program; some programs used multiple tool types and thus received multiple labels. Counting the number of labels for each type gave us a good impression of its commonness in Canadian STEM outreach (see results section below). Second, we identified specific measurements (i.e. questions) used within the two largest categories (i.e. student surveys and teacher/chaperone surveys). There was an enormous range of evaluation items, so we abandoned our initial goal to conduct a commonness count for each one; instead, we focused on developing a master list of all questions-types present, to illustrate the range (see results section). Carrying out these data collection and analysis processes also precipitated opportunities for further observation, explanation, and reflection for understanding and improving evaluation practice (see subsequent sections).

## Results

Our first direct output is a summary of evaluation methods for STEM outreach, as demonstrated by our sample of Canadian initiatives. Table 2 lists 11 different types of tools, including a description and a commonness percentage for our sample. The most common tools are participant surveys, chaperone surveys, and external metrics (e.g. simply counting the number of participants). Less common tools are facilitator observations, discussions, testimonials, host organization feedback, pre-surveys, written submissions, longitudinal tools, quizzes, and interviews. Note that the percentages add up to more than 100% because several initiatives use more than one evaluation tool and thus are counted under more than one row.

Table 2. Summary (or Toolbox) of Evaluation Methods for STEM Outreach Events

| **Tool** | **Description** | **Commonness in Our Sample** |
|---|---|---|
| Participant Survey | a survey given to participants (i.e. the target audience, such as students) after the outreach event | 35/68 (51.5%) |
| Teacher or Chaperone Survey | a survey given to a host teacher or chaperone (i.e. who can comment on the event more broadly and share observations of participants) after the outreach event | 23/68 (33.8%) |
| External Metrics | summary quantitative data collected indirectly (e.g. number of participants, location, demographics, repeat interest from same organization, etc.) | 13/68 (19.1%) |
| Facilitator Observation | unstructured observations and perceptions noted by the facilitator themselves during the outreach event or survey/debrief of facilitator after the outreach event | 8/68 (11.8%) |
| Discussion or Focus Group | a group conversation held with participants after the outreach event (includes sharing/talking circles) | 4/68 (5.9%) |
| Host Organization Feedback | formal or informal feedback requested from representatives of host organizations as a whole, rather than individual participants or teachers/chaperones | 3/68 (4.4%) |
| Separate Testimonial | request for participants or teachers/chaperones to write positive comments that the outreach initiative can use for public promotion (may also be part of a survey) | 3/68 (4.4%) |
| Pre-Survey (or Quiz) | a survey given to participants or chaperones before the outreach event (can be used to set up a before-and-after comparison or simply focus on pre-event coordination) | 3/68 (4.4%) |
| Longitudinal Tools | information collected long after the outreach event, through follow-up surveys or host organization feedback, to determine long-term impact | 2/68 (2.9%) |
| Written Submission | request for participants or teachers/chaperones to submit an unstructured written reflection or set of recommendations | 2/68 (2.9%) |
| Interview | individual interviews held with participants after the outreach event | 1/68 (1.5%) |

Before focusing on the most common tools (i.e. surveys) for the conclusion of this section, some additional details on some of the more complex tools are warranted. In our sample, external metrics included: number of events held, number of participants at each event, locations, list of classes/organizations the program was delivered to, whether audience classes /organizations were new or repeat, time required to prepare and deliver material, whether there was an intermediary partner organization, and demographic measures (i.e. gender breakdown, Indigenous participants, rural classes/organizations, socioeconomic status of classes /organizations, grade or age range). Questions on pre-surveys included: how the respondent found out about the program, what the respondent's expectations are, estimated knowledge about the topic (e.g. on a Likert scale), and quizzes assessing the respondent's knowledge of the topic. Most of these questions can be compared to a parallel question on a post-survey to assess the impact of the event. Some post-surveys asked respondents to estimate what they knew or thought before the event, instead of actually administering a separate pre-survey.

Our second direct output is question banks for the two most common evaluation tools: participant surveys (see Appendix A) and teacher or chaperone surveys (see Appendix B). These appendix tables include all the questions we encountered in our sample for each survey type, but with similar questions and phrases collapsed for simplicity. The tables indicate whether each question—in the sample—was asked as a yes/no question, a Likert-scale question, another single-choice question (i.e. pick one option from a list), a checklist question (i.e. pick as many options as desired from a list), a fill-in-the-blank question, or a more open qualitative question (e.g. short or long free-form answer).

In summary, developing the question banks revealed that evaluation surveys may include the following sections: general questions (e.g. name of event), demographic questions (e.g. age), questions about before the event (e.g. how did you find out about this event?), general questions about the event (e.g. what was good about this event?), specific questions about the event (e.g. how interactive was this event?), questions estimating the impact of the event (e.g. did this event make you more interested in the topic?), questions about specific parts of the event (e.g. how effective was Facilitator A?), and other/miscellaneous questions. The most common quantitative survey questions focused on: enjoyability, learning, interactivity, usefulness, and potential outcomes (e.g. whether the participant was more likely to take a course or pursue a career related to the topic). The most common qualitative survey questions emphasized: strengths of the event/program, weaknesses of the event/program, suggestions for the event/program, what was learned, and general feedback (i.e. a space for any other comments).

## Discussion

In pursuing our overall purpose of understanding how to do STEM outreach evaluation, we cover four themes in this section: commonness of evaluation, effectiveness of evaluation, timescale of evaluation, and target audiences for evaluation.

### *Commonness of Evaluation*

Our results measured the relative frequency of different evaluation tools in the sample (e.g. participant surveys were used in more than half of the cases and teacher or chaperone surveys were used in about a third of cases). The predominance of survey tools matches with the findings of previous literature, but other tools such as quizzes, follow-ups, and interviews appeared to get relatively more attention in the relevant literature than in the sample. It is possible that surveys are popular in the sample because they are perceived as the most appropriate, effective, or efficient tool. However, we would suggest that several explanatory factors are likely at play: surveys are simple and low-cost (both in terms of monetary costs and time costs); every case in the sample was funded by PromoScience and thus is responding to the same reporting requirements (see specifics in the Methods section) that may be most easily met by surveys; and the sample is predominantly outreach initiatives led by STEM practitioners (as opposed to the more substantial presence of social scientists and education researchers as authors in the literature), who may be more familiar with quantitative measures. So, ultimately, the commonness of tools may not necessarily be a good proxy for the effectiveness of those tools.

*Effectiveness of Evaluation*

What can we say, then, about evaluation effectiveness from our observations throughout the data collection and analysis process? From the literature, we know a few lessons about evaluation designs that are effective (e.g. see Bagiya, 2016): qualitative questions may be more useful than quantitative questions, longer-term follow-ups with host organizations or teachers can be instructive, evaluation strategies should be co-designed with communities or other stakeholders, and approaches to evaluation should be shared across practitioners. The emphasis on short-term quantitative tools we observed in the sample suggests that there may be greater opportunity for Canadian STEM outreach practitioners to use qualitative questions and partnered approaches. Our informal observation that many contacts were very enthusiastic about the work we were doing (i.e. asking to be informed about the results) suggests that there is desire to share best evaluation practices more widely in Canada. These findings have implications for improving STEM outreach approaches (see “Recommendations” section below), but we should acknowledge the constraints of the context—PromoScience programs run on short (i.e. 2- or 3-year) cycles with specific reporting requirements, under which long-term relationship building and more rigorous data collection are not the focus. Similar contextual limitations, the requirements of funding agencies, were observed in the related literature. While the system (i.e. NSERC PromoScience) has certainly facilitated a rich diversity of effective outreach initiatives, there may be implications for system-wide improvements in addition to improvements to individual outreach evaluation strategies. Surveys are likely to remain the most appropriate and efficient tool at our disposal, given constraints under the current context, but some of the possible improvements are very straightforward, like adding qualitative questions (and relationship-building can be paired with any evaluation method, including surveys).

*Time Scale of Evaluation*

The relevant literature suggested that previous studies of STEM outreach cases represent a range of approaches to time scale, from no consideration of longer-term impacts at all to on-the-day attempts at projecting potential lasting impacts to medium-term follow-ups several years after an outreach event—with most initiatives using only on-the-day measurements. Our sample showed a similar tendency towards on-the-day measurements, with perhaps less range than previously studied cases; only 2 programs in the sample used longitudinal tools. One program administered follow-up surveys to participants in subsequent years while another tracked the proportion of participants who complete their studies or find employment related to the topic. However, many of the programs primarily using on-the-day measurements still attempted to project long-term impacts; roughly half of the programs using participant surveys included a question asking whether the respondent was more likely to take a STEM course, pursue a STEM career, or take some other action related to the topic. The structure of PromoScience partially explains this observation; the grant length is 2-3 years (meaning potentially less opportunity for follow-up past the short term) and the reporting requirements (see Methods section) encourage projected impacts. It is also difficult to pursue longitudinal measures with student participants, because ethics requirements and School Board policies understandably preclude collecting student contact information from one-time events, so long-term approaches may be limited to teacher follow-ups or repeated offerings like summer camps (which may have repeat registrants). Ultimately, though, there may be appetite for longer-term measurements of STEM outreach impact in Canada, from both individual outreach programs and PromoScience more broadly, evidenced by the frequency of “projecting” questions.

*Target Audiences for Evaluation*

The cases reported in the literature as well as the cases in our sample covered a wide range of evaluation audiences, divided primarily into the categories of students (e.g. various grade and age ranges) and chaperones (e.g. teachers and parents). Similarly, both sources included a few examples of facilitator-focused tools, such as debriefs with facilitators or facilitator-filled forms to record observations of the participants. However, our sample seemed to focus more on using facilitators as an additional channel for information about participant experiences (e.g. asking them what they observed) rather than on the experiences of facilitators themselves, unlike a few of the facilitator-focused cases in the literature. So, there may be an opportunity for Canadian STEM outreach programs to consider another angle by engaging facilitators more directly in evaluation. After all, facilitators are often volunteers (who already have demanding full-time jobs in STEM) and it would be useful to have information on what motivates them to deliver outreach programs to begin with and what keeps them coming back; that is, how can we avoid burnout and keep outreach programs sustainable? Another important audience consideration is that Canadian STEM outreach programs funded by NSERC PromoScience, even more so than the cases in the broader literature, frequently wanted to track their reach among equity-deserving groups: women and girls, participants from rural or remote communities, Black participants, Indigenous participants, racialized participants, newcomers to Canada, participants from low-income families, neurodiverse participants, and participants with disabilities. One reason for this focus is likely because the PromoScience program's reporting requirements ask for these very metrics (see the Methods section), which is an excellent example of the positive influence that funding agencies can have over STEM outreach initiatives. Some initiatives are designed from the outset to engage specific equity-deserving groups, often developing entirely distinct approaches tailored to effectively reach and involve those communities. These examples lead to some of the recommendations in the next section.

## Recommendations for Outreach Practitioners

Given that many STEM outreach evaluations are designed using anecdotal experience rather than consulting available best practices (see background section) and given that there is a very wide range of evaluation tools and questions used in Canada (see results section), there is need for evidence-informed standardized evaluation tools to support evaluation design. In this section, we present two template evaluation tools—a survey for participants and a survey for participants or chaperones (e.g. teachers)—as a third major output of our study. We have developed these templates to incorporate a balance of the typical, the effective, the comprehensive, and the practical. Thus, our intent is for current or prospective outreach practitioners to use them as possible starting points for designing their own evaluations, but we also elaborate several considerations for alternate approaches and how to adjust the templates for different purposes.

The template surveys are included in Appendix C (survey for participants) and Appendix D (survey for teachers or chaperones), free to use under a CC-NC Creative Commons license (i.e. no attribution necessary, but not for commercial purposes, and you cannot portray the templates as your own creations). They include our interpretations of the most common questions on evaluation surveys synthesized into sensible layouts—i.e. not repetitive, not too long, grouping questions into categories—supported by principles of good survey design—i.e. avoiding biased language, intuitive question order, and especially supplementing quantitative question with qualitative comments (see Krosnick, 2017). We made sure to include some emphasis on future outcomes, given the emphasis we observed in the sample (and in the prior literature), but we focused on near-term outcomes, which are much easier to estimate. It bears repeating that we intend the templates to be starting points for evaluation design, supported by broader considerations such as (but not limited to) those in the following paragraphs.

When designing an evaluation survey, the number of questions, the specific questions, and the phrasing of questions in participant surveys must be adjusted as appropriate for a given participant group. For example, a survey for early grades should have fewer questions, less detailed demographic questions, and plainer question phrasing. Indeed, a key consideration is what will be appropriate for the intended audience. Recall that a range of evaluation tools is presented in Table 2 (i.e. we use the term “Toolbox” in that table deliberately) and banks of questions for survey tools are available in Appendices A and B. Note that the categorizations in these tables are somewhat subjective, based on our interpretations, so the toolbox and question banks are also meant to serve as a starting points for design, rather than exhaustive inventories.

Early in the evaluation design process, strongly consider the possibility of collaborating with representatives of the target outreach audience—if appropriate—to co-design the evaluation plan (or the outreach plan altogether). This consideration is particularly essential for equity-deserving groups, such as Indigenous communities (Conference Board of Canada, 2020; Henderson et al., 2015), but is relevant to all stakeholder and audience groups (Bagiya, 2016; Varner, 2014) because it helps to ensure that the evaluation (and the outreach event itself) will be valuable to the intended beneficiary. Such co-design processes can consider the templates, toolbox, and question banks presented in this paper, but should be open to all possibilities. Remember that our sample included several entirely unique approaches (i.e. very different from surveying participants and teachers or chaperones) in the context of outreach to certain equity-deserving groups. Related, for rural and remote communities, the question of whether to do in-person outreach (i.e. high event cost but good engagement with the few participants reached) or virtual outreach (i.e. low event cost and less genuine engagement but many participants reached) is challenging and may be effectively informed by collaborative discussions with representatives of the communities themselves before finalizing the outreach plan. Our own experience suggests that virtual outreach initiatives should enlist the assistance of on-site collaborators in administering evaluations, to keep the response rate high.

Evaluation implementation is as important as evaluation design. For instance, response rates will be higher for a survey if it does not require a response to every question; if a respondent does not want to answer a particular question but the survey forces them to (e.g. a virtual survey overusing “force response” or “mandatory answer” options), they will likely choose instead to simply not submit the entire survey. Our own experiences also suggest that response rates will be much higher if the survey is administered in the closing moments of the outreach event, rather than being sent out afterward, regardless of whether it is a paper or virtual survey. To ensure that sufficient attention and time is given to survey implementation, it should be included in event plans, administered immediately after the outreach activity, and assigned as a responsibility to a particular person (i.e. if there are multiple collaborators facilitating the outreach event).

Finally, it is important to consider whether an outreach evaluation is simply an internal evaluation for a certain outreach program or a broader research project (e.g. intent to publish the results, asking questions with implications beyond a single program’s design). The Tri-Council Policy Statement on Research Involving Humans (TCPS 2) is instructive for practitioners from institutions of higher education, clarifying that program evaluation for internal use does not count as “research” (see Article 2.5) but broader intents may meet the criteria triggering required review by the appropriate Research Ethics Board. Treating outreach evaluation as a broader research project brings more possibilities for impact and dissemination of the evaluation’s results, but ethics approval must be acquired before collecting data from human participants; the process can take several months at some institutions and thus adds time to the overall program timeline. Other ethics approvals (e.g. Indigenous community, school board) may also be required. However, ethical considerations are important even for evaluations that are not research projects (especially as they do not require oversight by a Research Ethics Board), so all practitioners interested in evaluation can benefit from guidelines like the TCPS 2. For example, see the ethics approval note at the beginning of Sami et al. (2020) as well as the notes on informed consent and confidentiality in Tsybulsky (2019, p. 571). Table 3 summarizes our recommendations for STEM outreach evaluation, building on Table 1.

Table 3. Updated Summary of General Lessons for Effective Design of
STEM Outreach Evaluation (Integrating Prior Literature with Our Results)

| |
|---|
| **Lessons for Design Process** |
| • set goals for the evaluation process (and the outreach initiative in general) proactively |
| • if possible, work with appropriate stakeholders to design the evaluation from the start (especially if equity-deserving groups are a target audience) |
| • if possible, leverage interdisciplinary collaboration—for example, between natural scientists familiar with outreach content and social scientists familiar with collecting data from human participants |
| • consider formative (mid-point) evaluations as well as summative (end-point) evaluations |
| • prioritize evidence over anecdotal experience in evaluation design |
| • consider evaluation focused on presenters/facilitators, not only participants/chaperones |
| • determine whether the evaluation will require ethics approval(s) |
| **Lessons for Specific Evaluation Tool (e.g. Surveys)** |
| • select a tool appropriate for the outreach and evaluation goals (see Table 2) |
| • consider a range of questions (see Appendices A and B) based on the outreach and evaluation goals |
| • keep the evaluation short and simple, especially if the event is fast-paced |
| • surveys should use 10 or fewer questions, ordered and grouped logically |
| • include qualitative questions (and consider prioritizing them over quantitative questions) |
| • acknowledge and counter the biases of question order and phrasing |
| • adjust the design to be appropriate for the target audience and age group |
| • use template surveys (see Appendices C and D) as a starting point, to see how the above five recommendations can be implemented |
| • but also consider revealed or observed measures over self-reported measures, if possible |
| • measure progress towards program goals; if those goals are longer-term outcomes, appropriate longer-term measures should follow the initial evaluation, if capacity allows |
| **Lessons for Evaluation Implementation** |
| • do not make individual evaluation questions mandatory |
| • included administering the evaluation in event plans (immediately after the outreach activity) and as the responsibility of a specific team member |

## Conclusion

This project set out to build knowledge on how to do STEM outreach evaluation by identifying and developing effective evaluation tools as well as outline considerations for their use. We have compiled guidance from the available literature (e.g. see Table 1) and, after surveying STEM outreach programs in Canada, have developed: a toolbox of approaches (e.g. see Table 2), question banks for use in participant and chaperone surveys (e.g. see Appendices A and B), starting-point templates for participant and chaperone surveys (e.g. see Appendices C and D), and general guidelines for STEM outreach evaluation (e.g. see Table 3). While there is a great range of approaches that can be taken to STEM outreach evaluation, as indicated by the resources we advance here, we acknowledge that contextual and resource constraints may limit what is possible for any given outreach program. However, we suspect most practitioners will find a few instances of "low-hanging fruit" among the possibilities (e.g. adding a qualitative question to an otherwise quantitative survey is a low-cost adjustment with high potential impact). Overall, we have set out to both highlight and supplement prior literature by adding a Canadian perspective, leveraging a large-N study, and synthesize comprehensive guidance on how to do STEM outreach evaluation by integrating our own results with the literature.

From a more personal perspective, we also learned some important lessons through the process of this collaboration, not just from its results. Most importantly, we witnessed the benefits of interdisciplinary collaboration for STEM outreach evaluation: in our case, the natural scientist provided the essential context and practice (i.e. experience with STEM outreach) and the social scientist provided insights around survey design, research ethics, and partnership building. Actually, Tran et al. (2024, p. 37) made a brief similar observation, suggesting that some of the common shortcomings in outreach evaluation design can be explained by limited social science representation on outreach teams (also see Varner 2014, p. 336). Essentially, STEM outreach evaluation is difficult for a single person, or a single-discipline team, to do effectively.

In terms of further work on this topic, we acknowledge that our objective was one of breadth rather than depth. That is, we were able to survey STEM outreach evaluation approaches broadly, and extract some important lessons from the commonness of different approaches, but we did not have the scope to assess the different approaches in depth. Future studies on STEM outreach evaluation could study, more directly and deeply, the potential effectiveness of different approaches, extending and responding to both Bagiya's (2016) research and our own. For example, interviews with practitioners who have experience with evaluation could reveal what tools and questions they have found to be the most valuable, providing evidence and examples for their impressions—or experimental approaches to evaluation could implement an outreach initiative using different evaluation tools across the iterations and settings of the program (especially including longitudinal tools over the long term), comparing the insights and evidence that come from the different tools.

Ultimately, we believe we have achieved our aim to bring together some comprehensive information on how to do STEM outreach evaluation, supported by specific starting-point tools for current and prospective practitioners. We hope these resources, insights, and tools will help contribute to enhanced public engagement for STEM practitioners, greater student interest in STEM careers, and increased STEM literacy across our society.

Appendix A: Question Bank for Surveys of Participants after STEM Outreach Events

| Question (Common Phrasing) | Response Types | | | | | |
|---|---|---|---|---|---|---|
| | yes or no | Likert scale | other single choice | check-list | fill a blank | open qualit-ative |
| General Information | | | | | | |
| date of event: | | | | | ✓ | |
| name of participant: | | | | | ✓ | |
| name of school or organization: | | | | | ✓ | |
| contact information [or parent’s/guardian’s name]: | | | | | ✓ | |
| what event [or program] did you participate in? | | | ✓ | | ✓ | |
| Demographic Information | | | | | | |
| name of home community: | | | | | ✓ | |
| postal code: | | | | | ✓ | |
| age [or grade]: | | | ✓ | | ✓ | |
| gender identity: | | | ✓ | | ✓ | |
| do you identify as Indigenous? | ✓ | | | | | |
| do you identify as a new Canadian? | ✓ | | | | | |
| Questions about before the Event | | | | | | |
| how did you find out about this program? | | | ✓ | | | ✓ |
| why did you sign up for this program? | | | | ✓ | | ✓ |
| General Questions about the Event (Overall Quality) | | | | | | |
| what was good about this event? | ✓ | | | | | ✓ |
| what was bad about this event? | ✓ | | | | | ✓ |
| what was the best part of this event? | | | | ✓ | | ✓ |
| what was the worst part of this event? | | | | | | ✓ |
| how enjoyable was this event? | ✓ | ✓ | | | | |
| how useful [or valuable] to you was this event? | | ✓ | | | | |
| what were the specific strengths of this event? | | | | ✓ | | |
| what were the specific weaknesses of this event? | | | | ✓ | | |
| would you recommend this event to others? | ✓ | ✓ | | | | |
| what was the overall quality of this event? | | ✓ | | | | |
| what suggestions do you have for improving this event in the future? | | | | | | ✓ |
| is there any other feedback you would like to share about this event? | | | | | | ✓ |
| Specific Questions about the Event | | | | | | |
| did this event meet your expectations? | ✓ | ✓ | | | | |
| how engaging [or interactive] was this event? | ✓ | ✓ | ✓ | | | |
| how inspiring was this event? | ✓ | ✓ | | | | |
| was this event appropriate for your age [or grade]? | ✓ | ✓ | | | | |
| how educational was this event? | ✓ | ✓ | | ✓ | | |
| was this event the appropriate length? | | ✓ | ✓ | | | |
| did this event move at the appropriate pace? | | ✓ | ✓ | | | |
| did this event deliver appropriate content? | | ✓ | | | | |
| what emotions did you feel during this event? | | | | ✓ | | ✓ |
| Questions Estimating the Impact of the Event | | | | | | |
| did this event help you to understand the topic better? | ✓ | ✓ | | | | ✓ |
| did this event make you more interested in the topic? | ✓ | ✓ | | | | ✓ |
| did this event make you more likely to take a future course on the topic? | ✓ | ✓ | | | | |
| did this event make you more likely to pursue a career in the topic? | ✓ | | | | | ✓ |
| did this event make you more likely to take action related to the topic? | ✓ | | | ✓ | | ✓ |
| did this event create new opportunities for you around the topic? | | | ✓ | | | |
| have you participated in this event [or another program event] before? | | | ✓ | ✓ | | |
| would you be interested to participate in this event again? | ✓ | | | | | |
| Questions about Specific Parts of the Event | | | | | | |
| how effective were the different activities? [separate question for each] | | ✓ | | | | |
| how effective were the facilitators? [separate question for each] | | ✓ | | | | |
| how effective was the organization [or support] for the event? | | ✓ | ✓ | ✓ | | |
| how effective [or appropriate] were the facilities for the event? | | ✓ | | | | |
| Other Questions | | | | | | |
| please provide any additional comments not covered above: | | | | | | ✓ |
| what is one thing you learned at this event? | | | | | ✓ | |
| [non-evaluative questions like "what is your favourite animal?"] | | | | | ✓ | |

Appendix B: Question Bank for Surveys of Teachers or Chaperones after STEM Outreach Events

| Question (Common Phrasing) | Response Types | | | | | |
|---|---|---|---|---|---|---|
| | yes or no | Likert scale | other single choice | check-list | fill a blank | open qualit-ative |
| General Information | | | | | | |
| date of event: | | | | | ✓ | |
| name of teacher [or chaperone]: | | | | | ✓ | |
| name of school or organization: | | | | | ✓ | |
| email address: | | | | | ✓ | |
| phone number: | | | | | ✓ | |
| how many participants were in the group? | | | ✓ | | ✓ | |
| what event [or program] did the group participate in? | | | ✓ | | ✓ | |
| Demographic Information | | | | | | |
| name of group's home community: | | | | | ✓ | |
| age [or grade] range: | | | ✓ | | ✓ | |
| gender breakdown: | | | ✓ | | ✓ | |
| is the group from a rural community? | ✓ | | | | | |
| how many members of the group are Indigenous? | | | ✓ | | ✓ | |
| how many members of the group are new Canadians? | | | ✓ | | ✓ | |
| how many members of the group are minorities? | | | ✓ | | ✓ | |
| how many members of the group come from low-income families? | | | ✓ | | ✓ | |
| how many members of the group have a disability? | | | ✓ | | ✓ | |
| General Questions about the Event (Overall Quality) | | | | | | |
| what was good about this event? | | | | | | ✓ |
| what was bad about this event? | ✓ | | | | | ✓ |
| what was the best part about this event? | | | ✓ | | | |
| what was the worst part about this event? | | | ✓ | | | |
| how enjoyable was this event for the group? | ✓ | ✓ | | | | |
| how useful [or valuable] was this event for the group? | | ✓ | | | | ✓ |
| how would you rate the overall experience for participants? | | ✓ | | | | |
| would you recommend this event to others? | | ✓ | ✓ | | | |
| what was the overall quality of this event? | | ✓ | | | | |
| what suggestions do you have for improving this event in the future? | | | | | | ✓ |
| is there any other feedback you would like to share about this event? | | | | | | ✓ |
| Specific Questions about the Event | | | | | | |
| did this event meet your expectations? | | | ✓ | | | ✓ |
| how engaging [or interactive] was this event for the group? | ✓ | ✓ | | | | |
| how inspiring was this event for the group? | | ✓ | | | | |
| was this event appropriate for the age [or grade] of the group? | | ✓ | | | | |
| was this event the appropriate length for the group? | | ✓ | | | | |
| was this event connected to the curriculum? | | ✓ | | | | |
| what was the quality of the event's different potential benefits: [separate question for each] | | ✓ | | | | |
| did different events [or programs] complement each other? | ✓ | | | | | |
| was cost of the event a potential barrier for this group? | ✓ | | | | | |
| Questions Estimating the Impact of the Event | | | | | | |
| was this event educational [or did it facilitate learning] for the group? | ✓ | | | | | |
| how much of the group do you think understands the topic better now? | | | ✓ | | ✓ | |
| how much of the group do you think is more interested in the topic now? | | | ✓ | | ✓ | |
| how much of the group do you think is more likely to take a course on the topic now? | | | ✓ | | ✓ | |
| how much of the group do you think is more likely to pursue a career in the topic now?[i] | | | ✓ | | ✓ | |
| will the group take action related to the topic now? | ✓ | | | | | |
| did this event create new opportunities for the group around the topic? | | ✓ | | | | ✓ |
| how many times have you attended this event [or program] with a group before? | | | ✓ | | ✓ | |
| would you be interested to attend this event [or program] with a group again? | ✓ | | | | | |
| Questions about Specific Parts of the Event | | | | | | |
| how effective were the different activities? [separate question for each] | | ✓ | | | | ✓ |
| how effective were the facilitators? [separate question for each] | | ✓ | | | | |
| how effective was the organization [or support] for the event? | | | | | | ✓ |
| how effective [or appropriate] were the facilities for the event? | | ✓ | | | | |
| Other Questions | | | | | | |
| please provide any additional comments not covered above: | | | | | | ✓ |
| do we have your permission to use your (anonymous) comments publicly? | ✓ | | | | ✓ | |
| do we have your permission to contact you for further feedback? | ✓ | | | | | |

[i] This phrasing was present in multiple surveys in the sample, but it is biased or "loaded" (i.e. it has a strong assumption of program impact across the group). There are ways to ask the same basic question with less biased language. For example, "In your view, how likely (on average) are participants in the group to try and learn more about [topic] because of this event?" (Likert scale) uses more neutral phrasing.

Appendix C: Sample Template Survey for Participants after STEM Outreach Events

Date: Name of School: Grade:

How fun was this event? (circle one)

(1) not fun at all (2) a little bit fun (3) somewhat fun (4) pretty fun (5) very fun

What was the best part of this event? ________________________________________

What was the worst part of this event? ________________________________________

How much did you learn about [topic] from this event? (circle one)

(1) nothing at all (2) a little bit (3) some (4) quite a bit (5) a lot

What is one thing you learned about [topic]? ________________________________________

Do you think you will try to learn more about [topic] because of this event? (circle one)

(1) no (2) possibly (3) maybe (4) probably (5) definitely

What will you do to learn more about [topic] now? ________________________________________

Please write anything else you would like to share with us: ____________________________________________________________

______________________________________________________________________________________________________________

______________________________________________________________________________________________________________

Appendix D: Sample Template Survey for Teachers or Chaperones after STEM Outreach Events

Date: Name of School: Your Name: Contact Info:

How many participants were in the group? ________________ What was the age or grade range of the group? ________________

How enjoyable was this event for the group? (circle one)

(1) not enjoyable at all (2) a little bit enjoyable (3) somewhat enjoyable (4) pretty enjoyable (5) very enjoyable

How valuable was this event for the group? (circle one)

(1) not valuable at all (2) a little bit valuable (3) somewhat valuable (4) pretty valuable (5) very valuable

Did this event meet your expectations as a teacher or chaperone? (circle one)

(1) not at all (2) met some (3) met (4) exceeded (5) far exceeded

In your view, how likely (on average) are participants in the group to try and learn more about [topic] because of this event?

(1) not likely at all (2) a little bit likely (3) somewhat likely (4) pretty likely (5) very likely

Please elaborate on why you have that view: ____________________________________________________________________________

______________________________________________________________________________________________________________

What was the best part of this event? ________________________________________________________________

What was the worst part of this event? ________________________________________________________________

Will the group take more action related to [topic] because of this event? If so, what? ___________________________________________

What suggestions do you have for improving this event in the future? ______________________________________________________

Would you be interested in attending an event with this organization again? ______________________________________________

---

Please provide any additional comments not covered above: ____________________________________________________________

______________________________________________________________________________________________

______________________________________________________________________________________________